\documentclass{aa}
\usepackage{graphicx}
\usepackage{color} 
\usepackage{natbib} 
\bibpunct{(}{)}{;}{a}{}{,} 

\newcommand {\HII}{\mbox{H\,{\sc ii}}}

\newcommand {\cmcub}{\mbox{cm$^{-3}$}}

\newcommand {\msol}{\mbox{M$_{\odot}$}}
\newcommand {\lsol}{\mbox{L$_{\odot}$}}

\begin{document}
   \title{Dust properties of the dark cloud IC\,5146}

   \subtitle{Submillimeter and NIR imaging}

   \author{C. Kramer\inst{1} \and
          J. Richer\inst{2} \and
          B. Mookerjea\inst{1} \and
          J. Alves\inst{3} \and
          C. Lada\inst{4}
          }

   \offprints{C. Kramer}
   \institute{I. Physikalisches Institut, Universit\"at zu K\"oln,
     Z\"ulpicher Stra\ss e 77, 50937 K\"oln, Germany\\
     \email{kramer@ph1.uni-koeln.de, bhaswati@ph1.uni-koeln.de}
   \and
     Mullard Radio Astronomy Observatory, Cavendish Laboratory,
     Madingley Road, Cambridge CB3 0HE, England\\
     \email{jsr@mrao.cam.ac.uk}
   \and
     European Southern Observatory, Karl-Schwarzschild-Strasse 2, 85748
     Garching, Germany\\
     \email{jalves@eso.org}
   \and
     Harvard-Smithsonian Center for Astrophysics, 60 Garden Street,
     Cambridge, MA 02138, USA\\
     \email{clada@cfa.harvard.edu}
             }

   \date{}
   
   \abstract{ 
     We present the results of a submillimeter dust continuum study of
     a molecular ridge in IC\,5146 carried out at 850\,$\mu$m and
     450\,$\mu$m with SCUBA on the James Clerk Maxell Telescope
     (JCMT). The mapped region is $\sim14'\times2\farcm5$ in size
     ($\sim2$\,pc$\times$0.3\,pc) and consists of at least four dense
     cores which are likely to be prestellar in nature. To study the
     dust properties of the ridge and its embedded cores, we combined
     the dust emission data with dust extinction data obtained by Lada
     et al. (1999) from the NIR colors of background giant stars. The
     ridge shows dust extinctions above $\sim10$\,mag, rising up to
     35\,mag in the cores.
     
     A map of dust temperatures, constructed from the continuum flux
     ratios, shows significant temperature gradients: we find
     temperatures of up to $\sim20\,$K in the outskirts and between
     the cores, and down to $\sim$10\,K in the cores themselves.  The
     cores themselves are almost isothermal, although their average
     temperatures vary between 10\,K and 18\,K.  We used the
     extinction data to derive in addition a map of the dust
     emissivity parametrized by
     $\kappa'=\kappa_{850}/\kappa_{\rm{V}}$.  The average value of
     $\kappa'$ agrees well with the canonical value of Mathis (1990).
     We find that $\kappa'$ increases by a factor of $\sim4$ from
     $\sim1.3\,10^{-5}$ to $\sim5\,10^{-5}$ when the dust temperature
     decreases from $\sim20$\,K to $\sim12$\,K. A Monte Carlo
     simulation shows that this change is significant with regard to
     the estimated calibration uncertainties.  This is consistent with
     models of dust evolution in prestellar cores by Ossenkopf \&
     Henning (1994) which predict that grain coagulation and the
     formation of ices on grain surfaces in the cold, dense cloud
     interiors lead to a significant increase of the $850\,\mu$m dust
     opacity. This interpretation is furthermore supported by the
     previous detection of gas-phase depletion of CO in one of the
     IC\,5146 cores (Kramer et al. 1999). Observations of dust fluxes
     at short wavelengths are however needed to verify this result.
     
} 

\maketitle
%
\section{Introduction}

Dense, cold cores embedded in molecular clouds are the progenitors of
protostellar cores which subsequently form low mass stars. Many
observational studies have recently focused on the determination of
the physical properties of these pre-protostellar or prestellar cores.
This is because the initial conditions just before the collapse phase
determine the properties of the protostar. There is consequently
considerable interest in measuring their masses and thermal balance.
Since cold molecular hydrogen cannot be detected in emission, trace
constituents of the cloud have to be studied, requiring knowledge of
their abundances.  Two approaches have generally been used. The first
is via the low-$J$ rotational emission lines of a rare CO isotopomer.
The major difficulty with this technique is that CO can freeze out
onto cold dust grains in the interiors of the prestellar cores thus
changing significantly the CO gas-phase abundance from its canonical
value of $\sim10^{-4}$
\citep[e.g.][]{bergin2002,tafalla2002,caselli1999,kramer1999} and also
thus changing the thermal balance of the cores \citep{goldsmith2001}.

Since dust is by far the most important coolant and constitutes about
1\% of the total mass, measuring the dust emission has proved popular
as another method. The recent advent of large, sensitive bolometer
arrays on millimeter and submillimeter telescopes now makes it
possible to measure the weak continuum emission even from cool clouds
showing no signs of star formation.  The surface regions of prestellar
cores are heated by the weak interstellar radiation field (ISRF) while
cosmic rays penetrate easily even the densest cores. The cores
therefore exhibit a small temperature gradient between the cold core
interiors and the warmer surfaces.  This argument was quantified by
early radiation transport models \cite[e.g.][]{leung1975} and refined
by recent calculations \citep[e.g.][]{zucconi2001,evans2001}. These
models show the dependence of the gradient on, for example, the grain
material and on the core geometry.  A comparison with observed fluxes
should therefore allow to discriminate between these factors.  Recent
observations have indeed shown that many prestellar cores exhibit a
temperature gradient while others appear to be isothermal
\citep[e.g.][]{ward-thompson2002}.

Another difficulty is that the grain opacity at submillimeter
wavelengths is poorly known and may vary due to grain evolution in the
dense, cold core interiors.  At wavelengths longer than about
250\,$\mu$m, the wavelength dependence of the emissivity is
customarily parametrized by a power-law function:
$\kappa_\lambda=\kappa_0(\lambda/\lambda_0)^{-\beta}$.  The dust
properties are thus characterized by two parameters, the emissivity
exponent $\beta$ and the opacity $\kappa_0$ at a given wavelength
$\lambda_0$.  In general, the opacity may not even follow a power
strict law, and so would require more than two parameters to describe
it; given the lack of accurate multi-frequency submillimeter data we
do not consider more complex models in this paper. The dust
coagulation model of \citet[][OH94 hence on]{ossenkopf1994} shows that
the formation of {\it dirty} ice mantles and grain coagulation
increase the opacity at long wavelengths
\citep[cf.][]{pollack1994,stognienko1995,henning1995}.

Some studies use the spectral energy distribution in the FIR and submm
regime to disentangle dust temperature and opacity
\citep[e.g.][]{ward-thompson2002}. These observations indicate a
constant spectral index of $\beta\sim2$.

Here, we try to tackle this task by combining optical extinction data,
obtained via NIR observations of background stars obtained by
\citet{clada1999}, with deep dust emission maps at 450 and
850\,$\mu$m, obtained with the SCUBA bolometer on the JCMT. There are
two key strengths of this technique.  First, the submillimeter flux
ratios in themselves are an excellent temperature diagnostic in cold
gas, if the dust index $\beta$ is nearly constant, because the
Raleigh-Jeans corrections are large at these frequencies
($h\nu/k_{B}=32$\,K at 450\,$\mu$m).  Second, the optical extinctions
are an accurate tracer of dust column densities independent of the
dust temperature.  Combining the three data sets therefore allows us
to determine at each position the dust temperature and the grain
opacity, averaged along the line of sight.

We observed an area of 2\,pc$\times$0.33\,pc of a filamentary cloud in
the IC\,5146 complex which consists of several dense starless
condensations.  This study thus covers a much larger region than the
previous work of \citet[][\,Paper I and II
henceforth]{kramer1998b,kramer1999} who studied one core of the
IC\,5146 filament.

\begin{figure*} 
\resizebox{\hsize}{!}{\includegraphics{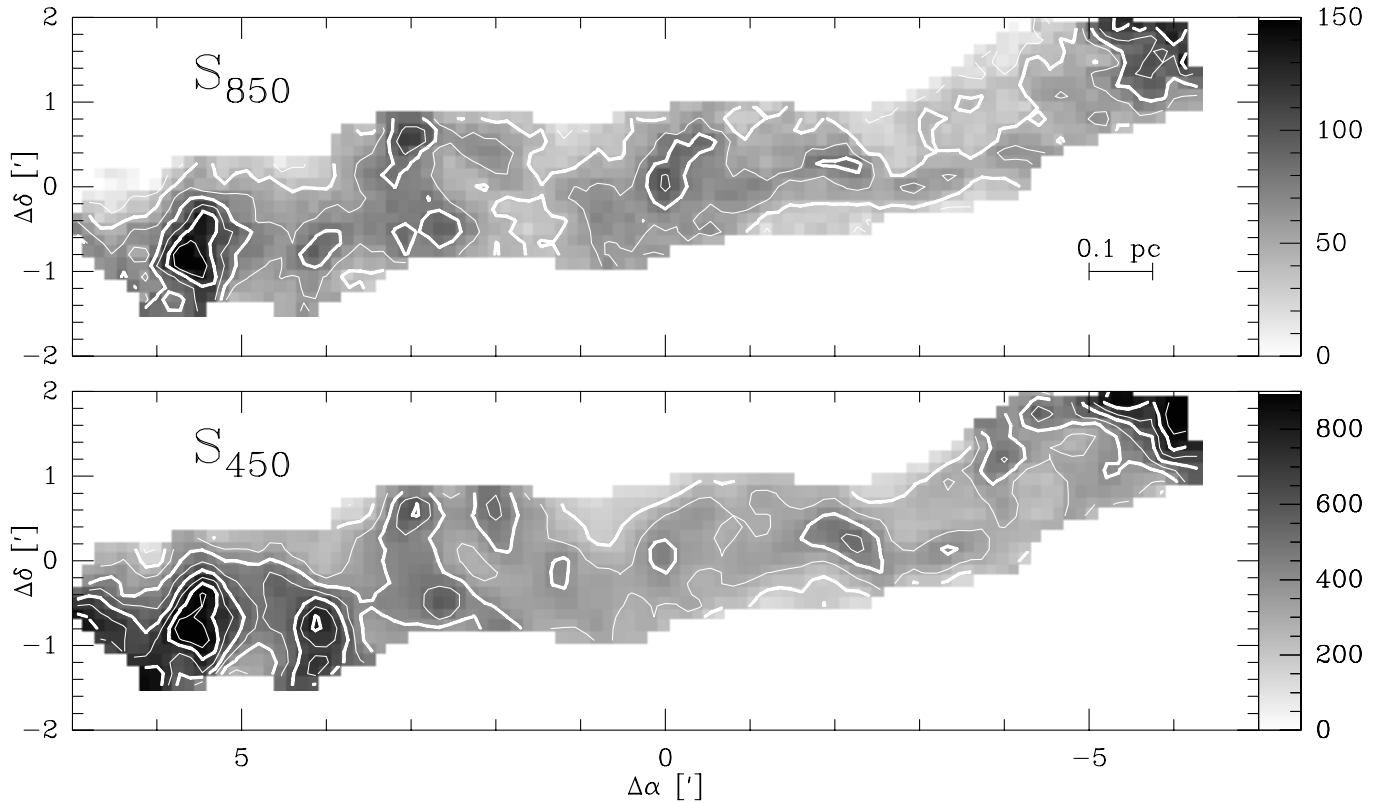}}
\caption{ \label{fig_map_850_450_16}
  Maps of flux densities at $850\,\mu$m and $450\,\mu$m. The
  $450\,\mu$m data are smoothed to the same resolution as the
  $850\,\mu$m data, both maps are at $16''$ resolution on a $8''$
  grid.  Contours of $S_{850}$ are 20 to 140 by 20\,mJy beam$^{-1}$.
  Contours of $S_{450}$ are 100 to 900 by 100\,mJy beam$^{-1}$.  The (0,0)
  position is $\alpha(2000)$ = $21\fh46\fm32\fs5$, $\delta(2000)$ =
  $47\fdg33\farcm55\farcs0$. }
\end{figure*}

\begin{figure*} 
\resizebox{\hsize}{!}{\includegraphics{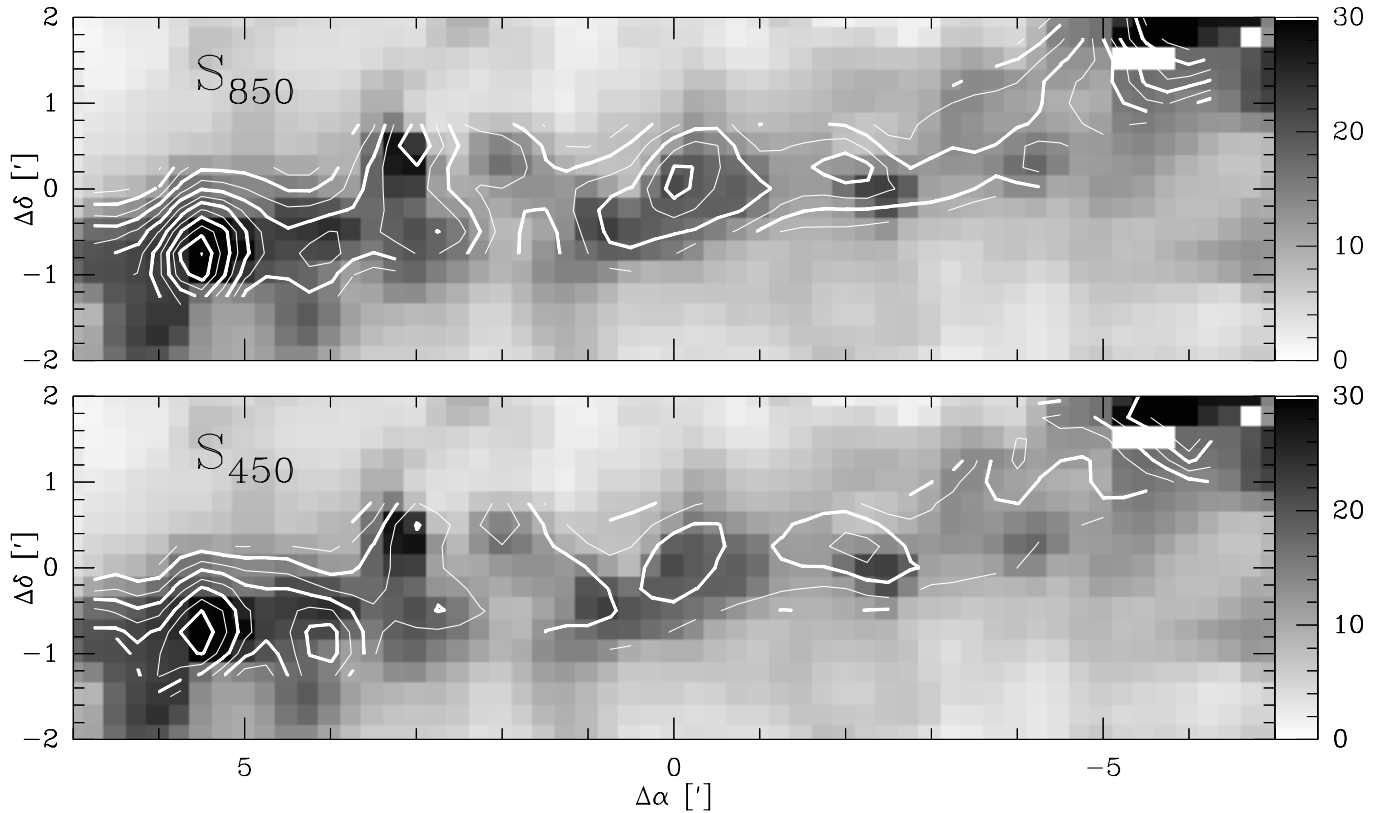}}
\caption{ \label{fig_map_850_450_av_30}
  Grey scale maps of optical extinctions overlayed with contours of
  flux densities at $850\,\mu$m and at $450\,\mu$m. Here, all three
  data sets were smoothed to a common resolution of $30''$ on a $15''$
  grid. Extinctions vary between 3 and 45\,mag.
  Contours of $S_{850}$ are 10 to 130 by 10\,mJy beam$^{-1}$.  Contours of
  $S_{450}$ are 80 to 900 by 80\,mJy beam$^{-1}$.  }
\end{figure*}

\section{Observations and data reduction}

\subsection{SCUBA data}

We used the Submillimeter Common User Bolometer Array
\citep[SCUBA,][]{holland1999} on the James Clerk Maxwell Telescope
(JCMT\footnote{The JCMT is operated by the JAC, Hawaii, on behalf of
  the UK PPARC, the Netherlands OSR, and the Canadian NRC.}) to map
IC\,5146. On September, 13th and 15th, 2000, we observed a
$14'\times2.5'$ region simultaneously at 850\,$\mu$m and 450\,$\mu$m.
The SCUBA array covers a hexagonal field of view of $2.5'$ with 91 and
37 pixels at 450 and 850\,$\mu$m respectively.  To create
fully-sampled maps, the `jiggle' mode was used, in which the telescope
beam is moved in a 64-position pattern by the secondary mirror,
spending 1\,s at each sample point while simultaneously chopping at
7.8\,Hz to remove atmospheric emission.  Since the filament is
elongated in right ascension and smaller than about $2'$ in
declination, we chopped in declination using the maximum throw of
$\pm2\farcm5$.  

Pointing, focus, and calibration checks were done on NGC\,7027 and
Uranus at regular intervals. The zenith opacity was derived from
simultaneous skydips at 850 and 450\,$\mu$m.  It stayed almost
constant at $\tau_{850}=0.18$ and $\tau_{850}=0.24$ during the first
and second night respectively, with a repeatability being better than
3\% at 850\,$\mu$m.  The half power beamwidths were measured on Uranus
to be $16''$ at 850\,$\mu$m and $10''$ at 450\,$\mu$m. The main beam
is almost Gaussian at 850\,$\mu$m.  At $450\,\mu$m the main beam
efficiency is $\sim50$\,\%, with the remaining power forming an error
beam around the main beam roughly $\sim30''$ in size.

\begin{figure*} 
\resizebox{\hsize}{!}{
  \includegraphics{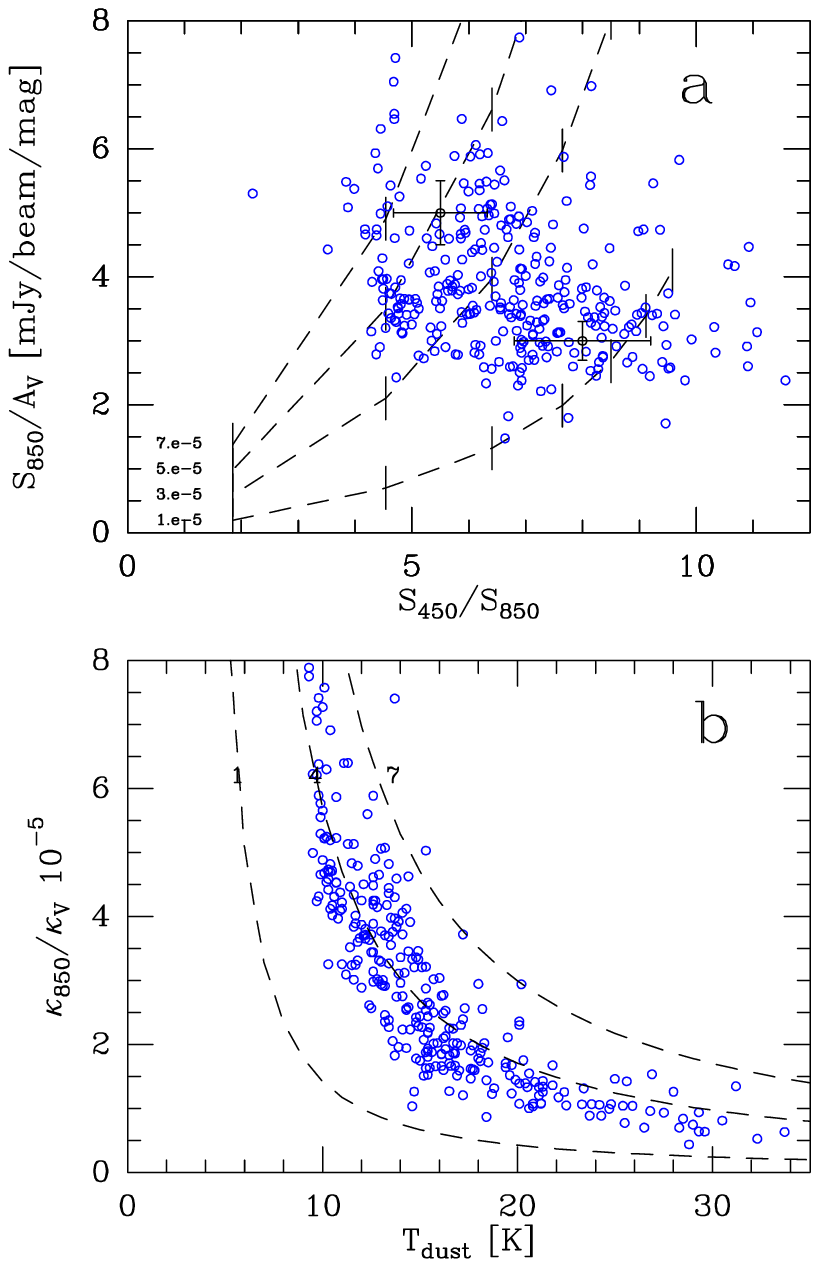} 
  \includegraphics{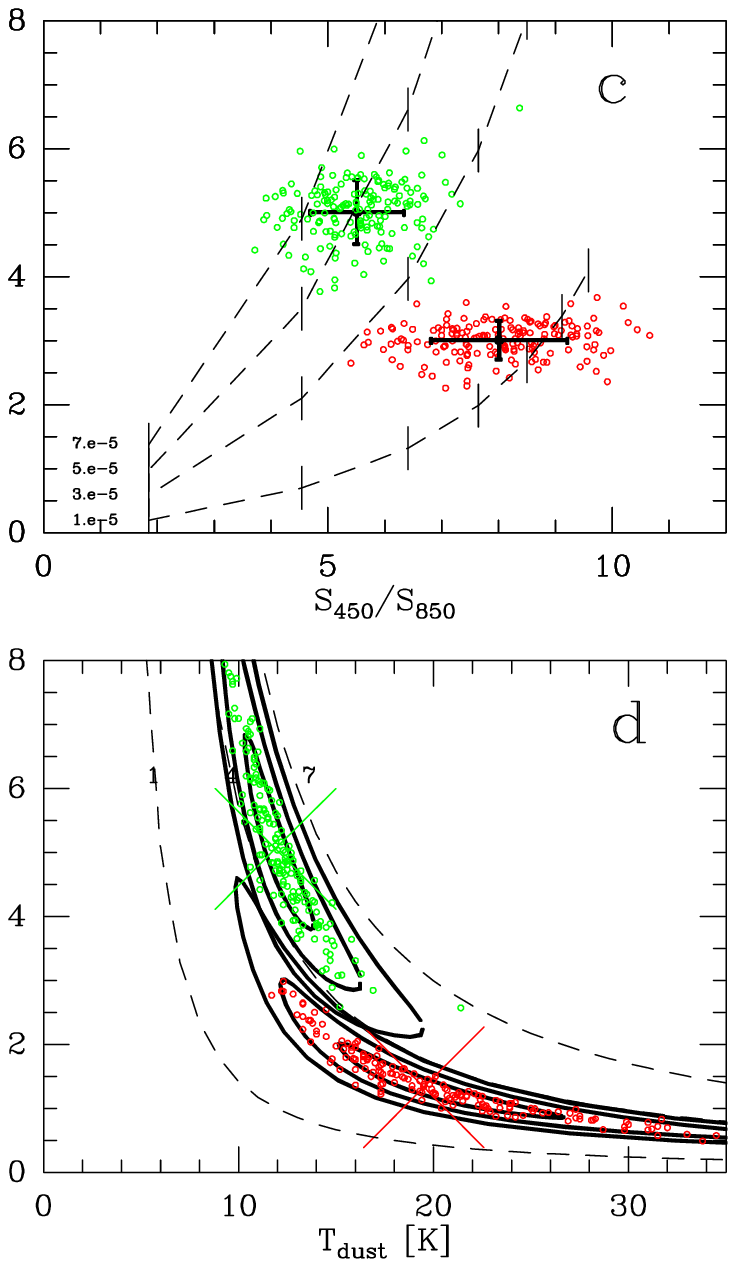} 
}
\caption{ \label{td_kprime} 
  {\bf (a)} Scatterplot of observed ratios $S_{850}/A_V$ versus
  $S_{450}/S_{850}$. Errorbars show the estimated calibration errors.
  Four dashed lines of constant kappa-ratio and varying dust
  temperature are also shown. These were calculated using Eqs.1,\,2
  assuming $\beta=2$.  Kappa-ratios range between $1\,10^{-5}$ and
  $7\,10^{-5}$ in steps of $2\,10^{-5}$.  Short perpendicular markers
  denote dust temperatures which range from 6 to 30\,K in steps of
  4\,K.  {\bf (b)} Scatterplot of $\kappa'$ versus $T_{\rm{dust}}$ for
  the mapped positions. Dust temperatures were derived from Eq.\,1
  assuming $\beta=2$. The $\kappa$ ratios were derived from Eq.\,2.
  The dashed lines delineate the temperature dependence of $\kappa'$
  for constant $S_{850}/A_V$=(1, 4, 7)\,mJy beam$^{-1}$ mag$^{-1}$
  from left to right.  {\bf (c)} Monte Carlo simulation of two data
  sets (see text for details). The widths of the normal distributions
  are the estimated calibration errors.  {\bf (d)} The resulting
  scatter plot of $\kappa'$ versus $T_{\rm{dust}}$.  The large cross
  marks the values derived from the mean values of $S_{850}/A_V$ and
  $S_{450}/S_{850}$. The three contours encircle the $1\sigma$,
  $2\sigma$, and $3\sigma$ areas.  }
\end{figure*}

We reduced the data primarily using the SCUBA User Reduction Facility
(SURF; \citet{jenness1998,jenness2002}) in the standard way.  The data
were corrected for individual bolometer gains (i.e.  flat-fielded),
for atmospheric extinction and correlated sky noise, and regridded and
calibrated to yield the final images in Jy\,beam$^{-1}$.  To improve
the image quality, we deconvolved both maps using our own CLEAN
algorithm to deconvolve the triple beam response of the observations:
this final step made only small changes to the maps because there is
little emission extended by more than 150 arcsec in declination.  At
450\,$\mu$m, our beam model is composed of three symmetric Gaussians
similar to the model fitted by \citet{hogerheijde2000}. We restored
the cleaned images back to the original resolutions. Next, we smoothed
the images to $30''$ resolution using a Gaussian convolving kernel.
And we resampled the data onto a common grid with $15''$ spacing. This
allows for a pixel-by-pixel comparison of the $450\,\mu$m and
$850\,\mu$m maps with the $A_{\rm{V}}$ data at the same resolution and
grid.

\subsection{Sources of errors}

The most important source of statistical error comes from fluctuations
in the atmospheric opacity.  At $450\,\mu$m, there are also possible
errors from changes in dish shape and thus in the beam pattern as the
dish temperature changes, and errors when deconvolving the error beam.
Finally, the {\it absolute\/} flux scales at 850 and 450\,$\mu$m are
derived from the flux of Uranus, and these are uncertain to $5-10$\%
\citep{sandell1994}. Since this error is entirely systematic, it
cannot account for any point to point variations in cloud properties,
only stretching monotonically the derived temperature or $\kappa'$
scales.

Uncertainties in the 450\,$\mu$m maps due to the error beam should be
small, because the CLEAN deconvolution included the error beam pattern
and because we are only interested in data at $30''$ resolution, which
includes most of the power in the error beam.  In addition, most of
the data were taken in cool nighttime conditions when the dish
temperature is stable.

The best estimate of the fluctuations in atmospheric transmission are
probably the opacities at 225\,GHz measured at the CSO every ten
minutes using a tipping radiometer \citep{archibald2000}. These
opacities varied very smoothly and by small amounts, indicating very
stable atmospheric conditions during our observations. We applied the
opacities at 850$\,\mu$m measured via SCUBA skydips to the data. The
scatter of these opacities, scaled to 225\,GHz \citep{archibald2002}
relative to the fitted opacities is less than $5$\% during the
11\,hours of the two night observations.  At the average CSO-zenith
opacity of 0.055 and typical airmass of 1.56, this translates into a
calibration uncertainty of $\sim10$\% at 450$\,\mu$m. Temporal
variations in opacity are expected to be slow, of the order of 10
minutes, so the error on close by parts of the maps will be much
smaller.

We thus estimate that the calibration error of the flux density ratio
$S_{450}/S_{850}$, observed simultaneously through the same
atmosphere, is $\sim15$\% and $\sim10\%$ for the ratio
$S_{850}/A_{\rm{V}}$ since the latter is dominated by the error of the
flux density.  In fact, the resulting dust temperatures of between 10
and 20\,K and their small scatter (Figs.\,4, 5) also indicate that
errors cannot be much larger.

\subsection{NIR data}

The NIR observations of IC\,5146 are described in
\citet{clada1999,clada1994}.  The determined color excess $E(H-K)$ is
proportional to dust column density.  Following convention, the dust
column density is represented in terms of optical extinctions $A_V$ by
converting the color excess at each position to the optical extinction
using the normal reddening law: $A_V=15.9 E(H-K)$ \citep{rieke1985}
with the ratio of total to selective extinction $R=A_V/E(B-V)=3.1$.
Although the resulting map accurately reflects the distribution of
dust column density and mass, it may not accurately predict the true
visual extinction through the cloud.  This is because grain growth in
cold clouds can alter the reddening law at wavelength $\ll 1 \mu$m
(e.g. Fig.2 in Mathis 1990).  Conversion to near-infrared extinctions
(e.g.  $A_K=1.78 E(H-K)$) would avoid this problem.  In this paper,
like in Papers I and II, we stick to $A_V$ and therefore compare the
dust absorption $\kappa_{850}$ in IC\,5146 to the absorption in the
NIR of general interstellar grains rather than to the actual value of
$\kappa_{\rm{V}}$ of grains in IC 5146.

\section{Properties of the dust}


\subsection{Maps}

The IC\,5146 complex of molecular clouds lies in Cygnus at a distance
of 460\,pc \citep{clada1999}. The part we analyzed is called Northern
Streamer \citep{dobashi1992,dobashi1993} for its filamentary
appearance which is strikingly visible in the map of optical
extinctions of \citet{clada1999}. We mapped at $850$ and $450\,\mu$m
the inner part of the ridge above about 10\,mag of optical extinction
covering an area of about $2\,$pc$\times0.3\,$pc. The resulting maps
of flux densities are shown in Fig.\,\ref{fig_map_850_450_16} at a
resolution of 0.04\,pc ($16''$). The filament has a clumpy appearance.
About nine cores are discernible at both wavelengths. Most prominent
is the bright clump at the eastern edge of the mapped region near
($5.3'$,$-0.8'$). Many clumps appear much more distinct at 850$\,\mu$m
than at 450$\,\mu$m, which is probably an indication that the dust is
cold, with the Rayleigh-Jeans correction suppressing the 450\,$\mu$m
emission.  This holds for example for the center core near
($0'$,$0'$). There are exceptions.  For example the clump near
($4.1'$,$-0.7'$) is much more distinct at 450$\,\mu$m than at
850$\,\mu$m which probably indicates a much higher fraction of warm
dust.

In Fig.\,\ref{fig_map_850_450_av_30} we show the same data set, but
smoothed to a common resolution of 0.067\,pc ($30''$) in comparison
with the optical extinctions at that resolution. Currently, the number
of detected background stars does not allow to create an $A_V$ map of
still higher spatial resolution. At least four cores show up at all
three wavelengths reaching peak extinctions between 20 and 35\,mag. An
exception is the core near ($0.8'$,$-0.5'$) which exhibits a peak in
extinctions of 28\,mag, while it does not show maxima in dust
emission. Almost no background stars are found near this position (see
Fig.\,1 in \citet{kramer1999zermatt}) but only one strongly reddened
star. This indicates high extinctions, but the position of the A$_V$
maximum may be shifted.

The mapped region shows no signs of star-forming activity. Two
protostellar cores associated with CO outflows and IRAS point sources
\citep[][]{dobashi1993,levreault1983}, the FU Ori source El1-12 with a
luminosity of 25\,\lsol\ and another much weaker protostar, lie just
outside of the SCUBA maps at offsets ($8.5'$,$-1.2'$) and
($-5.5'$,$2.47'$).  The \HII\ region S125 and its associated young
open cluster IC\,5146 lie more than $1^\circ$ to the east.

\subsection{Deriving the distribution of 
  dust properties over the surface of the map}

\subsubsection{Basic equations}

The dust temperature $T_{\rm{dust}}$ is determined by the ratio of
flux densities per beam at $850\,\mu$m and $450\,\mu$m which are
optically thin up to very high column densities:

\begin{equation}
  \label{eq_sratio}
  \frac{S_{450}}{S_{850}} =
  \Bigl(\frac{850}{450}\Bigr)^{3+\beta}
  \frac{\exp(17\,{\rm {K}}/T_{\rm{dust}})-1}{\exp(32\,{\rm{K}}/T_{\rm{dust}})-1}\,\, .
\end{equation}

At the low temperatures of about 10\,K prevalent in dark clouds and at
submillimeter wavelengths, deviations from the Rayleigh-Jeans law are
highly significant and so this ratio is a strong function of both the
dust temperature and the dust emissivity index $\beta$.  To derive the
dust temperature distribution over the surface of the cloud, we assume
$\beta=2$ as has generally been found for dust grains in molecular
clouds \citep{ward-thompson2002, johnstone1999, sandell1999,
  huard1999, visser1998, goldsmith1997} and is also consistent with
theory.  A relative error of the flux density ratio of 10\% would
result in an error of $T_{\rm{dust}}$ of 1.4\,K at 12\,K.

Next, we used the observed optical extinctions to derive the
distribution of the ratio of absorption and extinction coefficients
$\kappa'\equiv\kappa_{850}/\kappa_{V}$ over the surface of the cloud
via:

\begin{equation}
  \label{eq_s850av}
  \frac{S_{850}}{A_{\rm{V}}} = 
  B_{850}(T_{\rm{dust}}) \Omega_{850} \frac{1}{1.086}
  \frac{\kappa_{850}}{\kappa_{\rm{V}}}.
\end{equation}

$B_{850}(T_{\rm{dust}})$ is the Planck function at $850\,\mu$m and
$\Omega_{850}$ is the main beam solid angle. This analysis assumes
$\beta=2$, it may thus miss any variations of $\beta$, if present.
Dust flux densities in the FIR would be needed to analyze the
frequency dependence of the dust absorptivity in more detail. However,
we do derive variations of $\kappa'$. This approach is supported by
the dust models of e.g.  OH94 which indicate that grain evolution in
dense, cold prestellar cores leads to significant changes in
$\kappa_{850}$ but leaves $\beta$ rather constant.

\subsubsection{Scatterplots}

We use the three data sets $S_{850}$, $S_{450}$, $A_{\rm{V}}$ of the
full region at the common resolution of $30''$ and for each position
on a $15''$ grid.  Figure\,\ref{td_kprime}a shows a scatterplot of the
observed ratios $S_{850}/A_{\rm{V}}$ versus $S_{450}/S_{850}$. Mean
and rms values are $S_{850}/A_{\rm{V}}=4.0\pm1.5$\,(38\%) and
$S_{450}/S_{850}=6.7\pm1.6$\,(24\%). The scatter is thus significantly
larger than the estimated calibration errors of 10\% and 15\%
respectively. The reason may be variations of dust temperature and/or
of the $\kappa$-ratio $\kappa'$, assuming that $\beta$ stays constant.
Using the $A_{\rm{V}}$ data allows to disentangle these two
variations. We plot in Fig.\,\ref{td_kprime}a lines of constant
$\kappa'$ for dust temperatures in the range 6 to 30\,K using eqs. 1
and 2.  These lines clearly show that the observed data with its
uncertainty is inconsistent with a constant dust temperature and
kappa-ratio. High kappa-ratios are found for regions of low dust
temperatures. The resulting scatterplot of kappa-ratio versus dust
temperature (Fig.\,\ref{td_kprime}b) also shows this anti-correlation.
  
Care has to be taken when interpreting this plot, since the scatter of
points is influenced by the calibration errors and also by the
interdependency of the two equations used to derive the numbers for
$\kappa'$ and $T_{\rm{dust}}$. We therefore conducted a Monte Carlo
simulation to create two samples of data shown in
Fig.\,\ref{td_kprime}c. The rms scatter of these distributions equals
the estimated calibration errors. Their mean values lie at the low and
high end of the observed data but also well within its scatter
(Fig.\,\ref{td_kprime}a): $S_{450}/S_{850}=5.5\pm15\,\%$,
$S_{850}/A_{\rm{V}}=5\,$mJy beam$^{-1}$ mag$^{-1}$\,$\pm$10\%, and
$S_{450}/S_{850}=8.0\pm15\,\%$,
$S_{850}/A_{\rm{V}}=3\,$mJy beam$^{-1}$ mag$^{-1}$\,$\pm$10\%.  From these
distributions we derived again a scatterplot of $T_{\rm{dust}}$ and
$\kappa'$ (Fig.\,\ref{td_kprime}d). The two selected distributions are
significantly different at the $3\sigma$ level. We thus find a
significant drop of dust temperatures from 20\,K to about 12\,K while
the $\kappa$-ratio rises at the same time from $1.3\,10^{-5}$ to
$5\,10^{-5}$.


The average of the $\kappa$-ratio is $3.3\,10^{-5}$, comparing well to
the canonical value of $3.7\,10^{-5}$ derived from the extinction
curve of \citet{mathis1990} who uses the estimate of
\citet{hildebrand1983} for the FIR opacity:
\begin{eqnarray}
  \label{eq-mathis90}
 {\kappa_{\lambda}}/{\kappa_{\rm{V}}} & = & ({\lambda}/{250})^{-\beta} \,\,\,
 ({\kappa_{250}}/{\kappa_{\rm{J}}}) \,\,\,
 ({\kappa_{\rm{J}}}/{\kappa_{\rm{V}}}) \,\,\,\,\,\, \rm{and} \\
  \kappa_{\rm{V}}/\kappa_{\rm{J}} & = & 3.55, 
  \,\,\,\kappa_{250}/\kappa_{\rm{J}}=0.0015 \nonumber
\end{eqnarray}
with $\lambda=850\,\mu$m and $\beta=2$.  \\

\subsubsection{Dependence on assumed dust emissivity}

The absolute derived values of dust temperature and $\kappa'$ are
dependent on the assumed value $\beta=2$, but as long as $\beta$ is
constant, it cannot affect the derived point-to-point variations in
properties. The same is true if the absolute flux scales are in error.
For example, if $\beta=1.5$, the set of curves delineating constant
$\kappa'$ in Fig.\,\ref{td_kprime}a is shifted to the left.  The
resulting temperatures are increased. They range between 12 and
$\sim50$\,K and corresponding values for $\kappa'$ range between
$5\,10^{-5}$ and $0.5\,10^{-5}$.  The average value lies at
$1.7\,10^{-5}$, significantly lower than the canonical value.  Such
high dust temperatures are not expected for the weak interstellar FUV
field warming IC\,5146 (see next chapter).  We thus infer that a dust
emissivity index of $\beta=2$ is the more realistic value.  Of course,
if the assumed fluxes of Uranus (which set the absolute flux scales)
are in error, a different value of $beta$ may also be fully consistent
with the data.

\subsubsection{Residual emission at the OFF-positions}

The dust emission at the positions of the OFF-beam of the SCUBA jiggle
maps may be estimated from the large scale maps of optical extinctions
obtained by \citet{clada1999} and shown in
Fig.\,\ref{fig_map_850_450_av_30}.  To improve slightly, we used a
recently observed, new NIR map, covering the extended extinction of
IC\,5146 (unpublished observations by T.\,Huard, J.\,Alves, C.\,Lada).
The average optical extinction in the Northern area of the OFF-beam is
3.5\,mag, in the Southern area it is 2.8\,mag. For a rough estimate of
the corresponding dust emission at 850 and 450\,$\mu$m, we used eqs.1,
2, and assumed an average dust temperature of 20\,K and $\kappa$-ratio
of $1\,10^{-5}$ which is roughly appropriate for the inter-core
regions (cf.\,Fig.\,\ref{fig_profiles}). The resulting flux densities
are 8\,mJy beam$^{-1}$ at 850\,$\mu$m and 61\,mJy beam$^{-1}$ at 450\,$\mu$m
corresponding to about 6\% of the peak emission at both wavelengths.

The resulting error may be higher in the outskirts of the cores, where
dust emission is rather low and thus the relative contribution of
off-source emission may be high. This is e.g. the case at position
$(4.75'/-0.5')$ between cores \#2 and \#3, where flux densities drop
to $S_{850}=67$\,mJy beam$^{-1}$ and $S_{450}=517$\,mJy beam$^{-1}$.
The resulting dust temperature and $\kappa$-ratio at this position are
18.4\,K and $1.45\,10^{-5}$.  The average off-source contamination is
$\sim12$\,\% of the measured flux densities at both frequencies when
using the above rough estimate of the OFF-source flux densities.
However, the ratio $S_{450}/S_{850}$ is largely unaffected and thus is
the dust temperature determination.

We did not correct for this error which leads to a slight but
systematic underestimate of fluxes.

\subsubsection{Maps of $T_{\rm{dust}}$ and $\kappa'$}

The resulting maps of $T_{\rm{dust}}$ and $\kappa'$ are shown in
Fig.\,\ref{fig_map_td_nh2_k}.  Dust temperatures are low and lie
between $\sim10$ and 20\,K for the bulk of the filament.  Still higher
temperatures are reached at a few positions at the edges of the ridge
but are probably insignificant as discussed above.  The average and
rms are $\langle T_{\rm{dust}}\rangle=16.5\pm8$\,K.  In the center
region, dust temperatures are low at about 10\,K and there is only
little variation. To the east of the center, the map of dust
temperatures shows vertical stripes: temperatures are low at the
positions of clumps \#3 and \#2, while the temperature rises to above
$\sim20$\,K in the interclump regions and at the eastern edge of clump
\#2.  The western part of the map is less well characterized.
%
%
%

\subsubsection{Discussion}

Our finding is consistent with $\kappa_V$ and $\beta$ being constant,
while the absorption coefficient $\kappa_{850}$ is rising with
dropping temperatures.

\begin{figure*} 
\resizebox{\hsize}{!}{\includegraphics{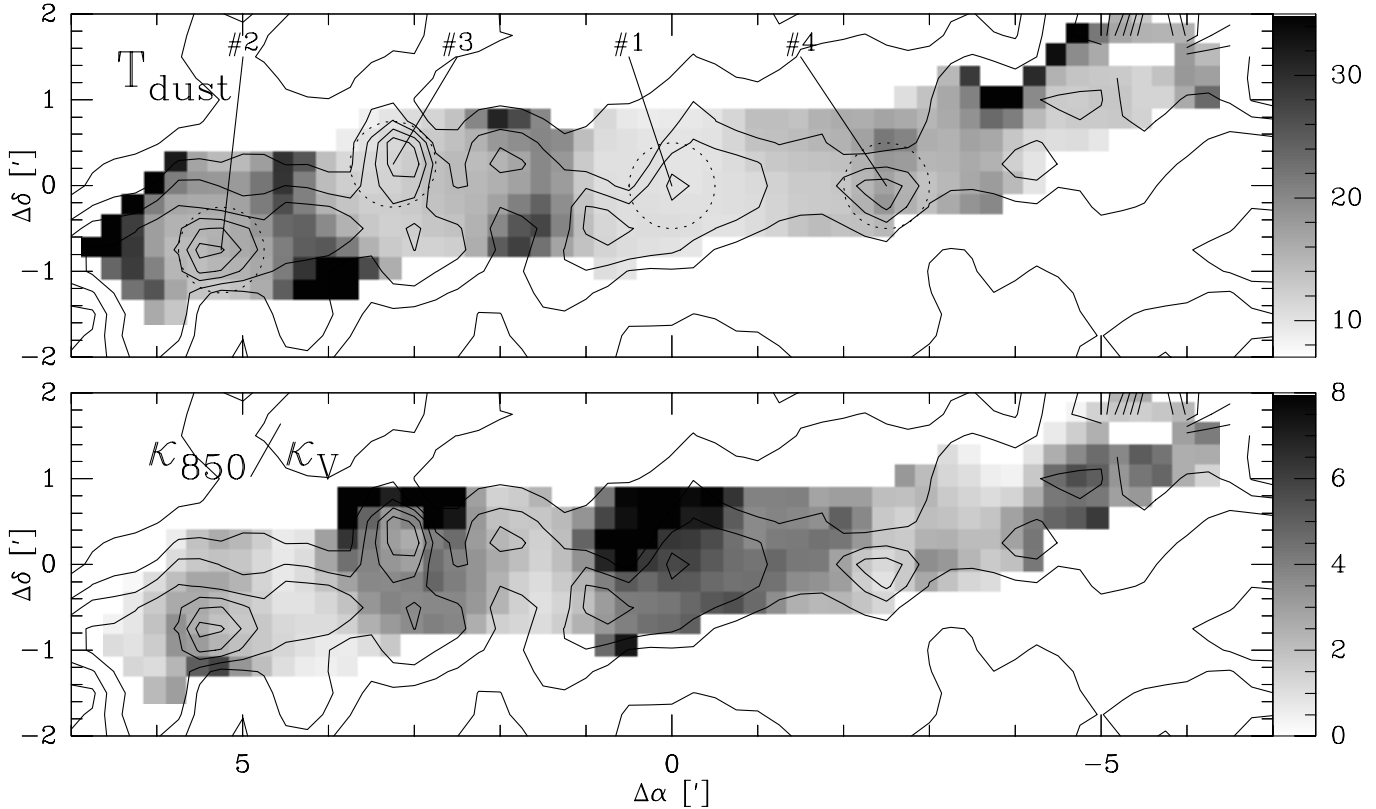}}
\caption{ \label{fig_map_td_nh2_k}
  Grey scale maps of dust temperature $T_{\rm{dust}}$
  and the ratio $\kappa'\equiv\kappa_{850}/\kappa_V$ together
  with contours of optical extinctions. Contours of $A_V$ start at
  5\,mag in steps of 5\,mag to 45\,mag. The spatial resolution of all
  data is $30''$.  Arrows denote the center positions of four
  prestellar cores which are studied in more detail.  Circles of
  0.067\,pc ($30''$) radii are centered at the core positions and
  indicate the region used to calculate core masses.  The dust
  temperature derived from the ratio of flux densities at 450 and
  850$\,\mu$m is in Kelvin.  
  The ratio $\kappa'$ was determined using in addition the optical
  extinction data is given in units of $10^{-5}$.  }
\end{figure*}

The colder regions are in general associated with regions of high
column densities as is seen in the map of $T_{\rm{dust}}$
(Fig.\,\ref{fig_map_td_nh2_k}) and in the radial profiles of the four
cores discussed in the next chapter (Fig.\,\ref{fig_profiles}).
Assuming that high column densities correlate with high volume
densities, grain coagulation may be of importance.
In addition, gas-phase molecules may freeze out on dust grains at low
temperatures, altering their emissivities.  \citet{sandford1993} e.g.
found a sublimation temperature of CO onto CO-ice of 17\,K from
laboratory experiments under interstellar conditions.  In Paper II, we
indeed found evidence for depletion of the CO gas-phase abundance in
one of the IC\,5146 cores, which we attributed to freeze-out on grain
surfaces.

This view is supported by the results of the dust model of OH94 who
show that coagulation and formation of ices on grains in the dense and
cold cloud interiors leads to an increase of long wavelength
absorption. Depending on parameters like the density or ice thickness,
OH94 predict enhancements by upto a factor of 3.4 (Table\,2 in OH94)
while we have observed an increase by about a factor of 3.8
(Fig.\,\ref{td_kprime}).
Grain coagulation has a larger influence than formation of ices which
contributes only a factor of 1.7. Dust opacity spectra of icy grains
displayed in OH94 (Fig.\,5b,c) show the rise in long wavelength
opacity with coagulation. These calculations also indicate that the
emissivity index $\beta$ stays almost constant for ice covered grains.
  
  Furthermore, Fig.\,5 in OH94 shows that the slopes of the dust
  opacity spectra stay rather constant in the NIR while the absolute
  opacities in the NIR rise only slightly when coagulation becomes
  important. The constancy of the slopes is consistent with the NIR
  extinctions law analyzed by \citet{clada1994} who found that the
  ratio of color excess $E(J-H)/E(H-K)$ stays constant upto optical
  extinctions of 30\,mag.

\subsection{Four prestellar cores in more detail}
\label{sec-radial-profiles}

\begin{figure*} 
\resizebox{\hsize}{!}{
  \includegraphics{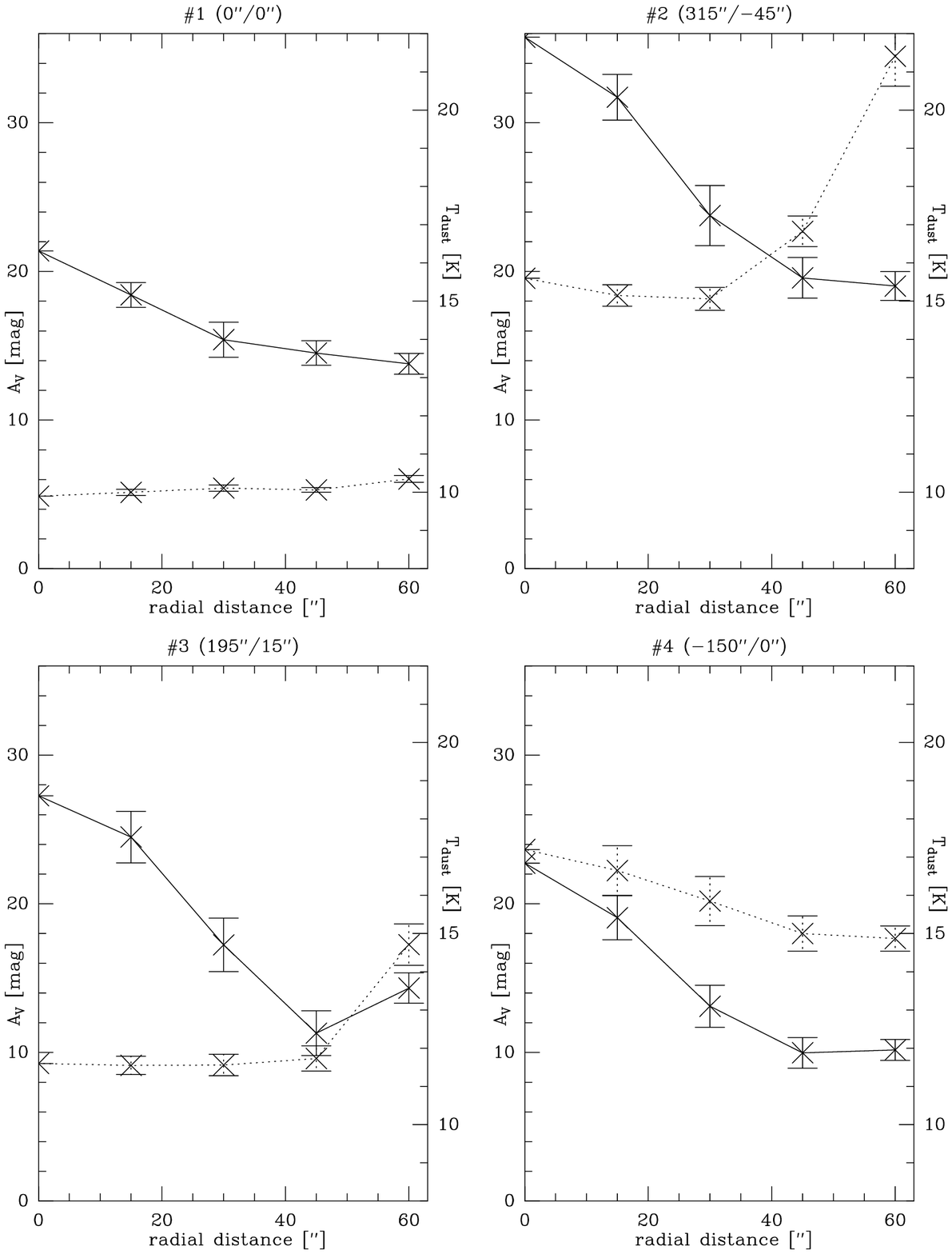}
  \includegraphics{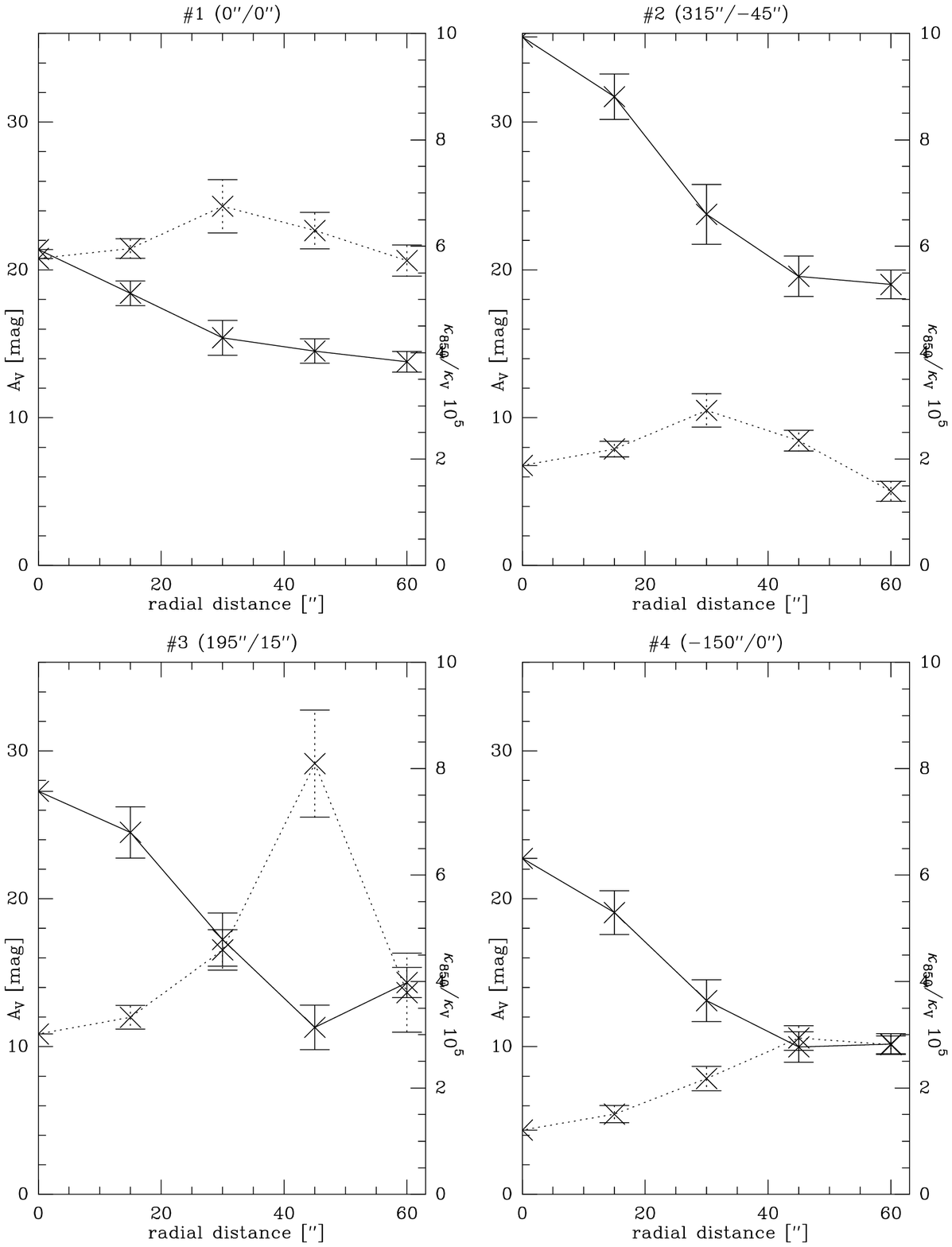}
}
\caption{ 
  \label{fig_profiles} 
  The left $2\times2$ plots show radial profiles of dust temperature
  (dotted lines) and optical extinction (drawn lines) for four cores
  showing peak extinctions greater than 20\,mag.  The right $2\times2$
  plots show radial profiles of $\kappa'$ (dotted lines) versus
  optical extinction (drawn lines) for the four cores.  The error bars
  show the rms error within each annulus, which includes noise and any
  deviations from circular symmetry in the cores.
}
\end{figure*}

The positions of four embedded cores are shown in the map of optical
extinctions (Fig.\,\ref{fig_map_td_nh2_k}). Center extinctions rise
above 20\,mag and all cores show some central symmetry. We therefore
derived radially averaged profiles of optical extinction, dust
temperature and $\kappa'$ (Fig.\,\ref{fig_profiles}).  Radially
averaged extinctions drop by about 50\% at a distance of $1'$
corresponding to 0.13\,pc.  Mean radial values are listed in
Table\,\ref{tab_cores}.

The peak in $A_{\rm{V}}$ near ($45''$,$-30''$) was excluded since it
exhibits an east-west orientated gradient in dust temperature and
$\kappa'$, but no central symmetry.

\begin{table*}
\caption[]{
Columns 3 and 4 give average dust temperatures and emissivities of the four
cores and of the total region mapped. Core averages are derived from
the radial profiles for $R\le60''$.  Average values were weighted 
with the number of pixels per annulus. The last line shows average 
values of the total mapped region covering all pixels. \\
Columns 5 to 7 give core masses within a radius of
$30''$ and masses of the total region mapped at 850 and 450\,$\mu$m.
$M_{\rm{Av}}$ is the mass calculated from the $A_{\rm{V}}$
data. $M_{\rm{canon}}$ is calculated independently from the $S_{850}$ flux
densities using $T_{\rm{dust}}=12\,$K and 
$\kappa_{850}=0.01$\,cm$^{2}$g$^{-1}$.  $M_{\rm{core}}$
is the mass calculated using the derived radial profiles of
$T_{\rm{dust}}$ and $\kappa'$. Mass ratio are presented in Col.\,8 and 9:
$R_1=M_{\rm{canon}}/M_{\rm{Av}}$, $R_2=M_{\rm{core}}/M_{\rm{Av}}$. 
}
\label{tab_cores}
\begin{tabular}{lcccccccc}
\noalign{\smallskip} \hline \noalign{\smallskip}
(1) & (2) & (3) & (4) & (5) & (6) & (7) & (8) & (9) \\

no. & $\Delta\alpha$/$\Delta\delta$ & $\langle T_{\rm{dust}} \rangle$ 
  & $\langle \kappa' \rangle$ 
  & $M_{\rm{Av}}$ & $M_{\rm{canon}}$ & $M_{\rm{core}}$ & $R_1$ & $R_2$  \\

    & & K & $10^{-5}$ & [\msol] & [\msol] & [\msol] & & \\
\hline \noalign{\smallskip}
\#1 & $0''$/$0''$     & $10.2\pm0.1$ & $6.1\pm0.3$ 
    & 4.12 & 4.43 & 3.56 & 1.07 & 0.86 \\ 

\#2 & $315''$/$-45''$ & $18.0\pm0.5$ & $2.1\pm0.2$ 
    & 6.66 & 6.08 & 5.66 & 0.91 & 0.85 \\ 

\#3 & $195''$/$15''$  & $12.6\pm0.4$ & $5.5\pm0.7$ 
    & 4.97 & 4.37 & 4.19 & 0.88 & 0.84 \\ 

\#4 & $-150''$/$0''$  & $15.4\pm0.5$ & $2.6\pm0.2$ 
    & 3.85 & 2.79 & 3.19 & 0.72 & 0.83 \\ 

sum &                 &              &
    &19.6  &17.7  &16.6  & 0.90 & 0.85 \\

total &             & $16.5\pm8$   & $3.3\pm2.6$ 
    &106.5 &124.4 & -    & -    & -    \\
\noalign{\smallskip} \hline 
\end{tabular} 
\end{table*} 

\subsubsection{Dust temperatures}


The average temperatures of the four cores differ significantly: they
range between 10\,K and 18\,K (Table\,\ref{tab_cores}). Mean
temperatures of cores \#1 and \#3 are only 10 and 12\,K,  much lower
than the globally averaged temperature of 16.5\,K. These findings
correspond well with typical dust and gas temperatures of starless
molecular cores observed previously. The survey of ammonia in dense
cloud cores by \citet{benson1989} shows typical gas temperatures of
between 10 and 15\,K. The observational derivation of dust
temperatures is more difficult.  Often, a canonical dust temperature
of 12\,K is used for quiescent dark clouds
\citep{visser2001}. 
%
Dust- and gas temperatures couple only at high densities of more than
$10^5$\,\cmcub\ \citep{goldsmith2001,kruegel1984}.

Cores \#1 and \#4 appear to be almost isothermal. They show an almost
constant temperature profile of 10\,K and 16\,K, respectively. The
other two cores show a constant inner temperature of 15\,K (core \#2)
and 12\,K (core \#3) for $R<40''$, and a rise of the temperatures by 7
and 3\,K respectively at greater radii which trace the 
interclump medium when the radially averaged optical extinctions
stay constant.  This rise is also prominent in the map of dust
temperatures (Fig.\,\ref{fig_map_td_nh2_k}), in particular for the
eastern most core \#2.

For the center region of the IC\,5146 filament, we have in addition
obtained a small map ($0.3\,$pc$\times0.3$\,pc, $2.2'\times2.2'$) of
1.2\,mm flux densities which had been presented in Paper\,I.  We
derived $T_{\rm{dust}}$ from the observed $S_{450}/S_{850}$ ratio and
$\kappa'$ from the independently observed $S_{1200}/A_V$ ratio
(assuming again $\beta=2$). The resulting dust temperatures vary only
between 9 and 14\,K.  There is again a systematic anti-correlation
between $T_{\rm{dust}}$ and $\kappa'$: the dust temperature varies
from 9\,K and $\kappa'=8\,10^{-5}$ to 14\,K and $\kappa'=1\,10^{-5}$.
The dust temperature drops slightly from an average of $\sim12$K at
10\,mag optical extinction to 10\,K at 20\,mag. This gradient is
shallower than the gradient derived solely from $S_{1200}/A_V$ in
Paper I, where a constant, canonical dust emissivity had been assumed.

A natural explanation for inwardly decreasing temperature gradients is
that these cores are prestellar and only externally heated by the weak
ambient ISRF. This leads to cool, dense dust in the core interiors
surrounded by outer, less dense layers of warmer dust.  
%
%
The FUV field impinging on IC\,5146 is probably slightly enhanced due
to the small IC5146 star cluster \citep{wilking1984} which lies at a
distance of $\sim1^\circ$ to the east.  Assuming that BD$+46.3474$, a
B0V star and the most massive member of the cluster, is the only
source of FUV photons, that the star is a black body at the effective
temperature of 30900\,K corresponding to its spectral type
\citep{panagia1973}, and that its radiation is not diluted by material
along the way, we derive a flux of $\chi=3$ in units of the average
interstellar radiation field at a distance of 8.1\,pc.a \footnote{ The
  integrated flux $\chi$ of the FUV field is given here in units of
  $\Phi_{\rm{D}}=2.7\,10^{-3}$ erg cm$^{-2}$ s$^{-1}$
  \citep{draine1996}. This unit is a factor of 1.71 stronger than the
  Habing field $G_0$ \citep{habing1968}.  }

\citet{ward-thompson2002} observed 18 prestellar cores using ISOPHOT
data at 170 and $200\,\mu$m.  From the flux ratio, assuming $\beta=2$,
they also find similarly low dust temperatures of between $\sim10$ and
20\,K. About half of their sources show a uniform temperature
distribution while the other half shows an inwardly decreasing
temperature profile. 

We find low l.o.s. averaged dust temperatures of $\sim$10\,K in some
cloud cores. This corresponds well with models of \citet{mathis1983}
who find that deep inside clouds, both silicate and graphite grains
attain local temperatures between 5 and 7\,K.  We have found
indications for increased opacities in the cloud interiors due to
grain coagulation and formation of ices. This increased emissivity may
lead to more efficient cooling of the dust grains, thus further
lowering the dust temperatures.  

Recent models by \citet{zucconi2001} and \citet{evans2001} quantify
the temperature gradient expected for prestellar cores illuminated
only by the external average ISRF.  These models show that local
temperatures vary between $\sim$8\,K in the interior and 14\,K at the
surface. An increase of the ISRF by a factor of 2 would lead to an
increased dust temperature by only 15-20\%. 

\subsubsection{Dust emissivity}

The average dust temperatures and $\kappa'$ values
(Table\,\ref{tab_cores}) are anticorrelated, as had already been found
when analyzing all the mapped positions (Fig.\,\ref{td_kprime}). Cores
\#2 and \#4 show a low average ratio of $2\,10^{-5}$ (and at the same
time relatively high temperatures) while cores \#1 and \#3 show a high
ratio of 5 and $6\,10^{-5}$ respectively (and relatively low
temperatures). The radial profiles (Fig.\,\ref{fig_profiles}) are
mostly flat (\#1, \#2, \#4). Only core \#3 shows a large radial
variation of the kappa-ratio between 3 and $8\,10^{-5}$.

\subsubsection{Masses}

We calculated the masses of the prestellar cores
(Fig.\,\ref{fig_map_td_nh2_k}, Table\,\ref{tab_cores}) by integrating
over a disk with a radius of 0.067\,pc ($30''$), i.e. covering the
inner region of the radial profiles shown in Fig.\,\ref{fig_profiles}.
We use three different methods to derive core masses. {\bf (1.)}
Optical extinctions were used to derive masses $M_{\rm{Av}}$ via the
canonical ratio of
$N($H$_2)$/$A_{\rm{V}}=9.36\,10^{20}$\,cm$^{-2}$mag$^{-1}$ observed by
\citet{bohlin1978} in the diffuse interstellar medium:
$M_{\rm{Av}}=N({\rm{H}}_2)\, A\, \mu\, m_{\rm{H}}$ with the area $A$
and the mean particle mass $\mu=2.33$ in units of the hydrogen atom
mass $m_{\rm{H}}$.  {\bf (2.)}  The flux density at $850\,\mu$m was
also used to independently derive the total mass $M_{\rm{canon}} =
{F_{850} d^2}/{(\kappa_{850} B_{850}(T_{\rm{dust}}))}$, where $d$ is
the distance, and $F_{850}=S_{850}/\Omega_{850}\times A$ is the
integrated flux. Here, we used canonical values for the dust
temperature and $\kappa$: $T_{\rm{dust}}=12$\,K and
$\kappa_{850}=0.01$cm$^{2}$g$^{-1}$.  The masses derived from dust
emission are a sensitive function of dust temperatures. For example,
assuming a temperature of 10\,K would result in an overestimate by a
factor of 2.1 relative to the mass at 15\,K temperature. Assuming on
the other hand a temperature of 20\,K would lead to an underestimate
by a factor of 0.6 relative to the mass at 15\,K.
{\bf (3.)} We therefore employed a third method, using the radial
profiles of $T_{\rm{dust}}$ (Fig.\,\ref{fig_profiles}) and $\kappa'$
to derive masses $M_{\rm{core}}$ \footnote{To calculate $\kappa_{850}$
  from the $\kappa'$ profile, we assumed a constant extinction
  coefficient in the V band of
  $\kappa_{\rm{V}}=273.6\,$cm$^{2}$g$^{-1}$. This value was in turn
  derived using the canonical values of \citet{mathis1990}
  (Eq.\,\ref{eq-mathis90}) and a mean $\kappa_{850} =
  0.01$cm$^{2}$g$^{-1}$.}.

Core masses lie between 3\,\msol and 7\,\msol, thus constituting only
about 20\% of the total mass. The masses derived by the three methods
agree to within 30\%.  This excellent agreement is surprising, given
that the masses $M_{\rm{Av}}$ and $M_{\rm{canon}}$ were derived from
two independent data sets i.e. the $A_V$ data on the one hand and the
submillimetric flux densities on the other hand and given the
observational errors of both methods. The agreement shows that the
observational errors are still smaller and the underlying assumptions
are valid to within that accuracy.

When comparing $M_{\rm{Av}}$ with core masses $M_{\rm{core}}$ derived
from the profiles of dust temperature and $\kappa'$, we find that the
latter are smaller by an almost constant factor of $0.85\pm0.01$. The
scatter is significantly smaller than the scatter of
$M_{\rm{canon}}$/$M_{\rm{Av}}$. The main factor contributing to the
improved mass estimate is probably the determined dust temperature.  A
correction of the canonical conversion factor N(H$_2$)/$A_V$ or,
alternatively, of the extinction coefficient $\kappa_V$ by only 15\%
would be needed to explain the constant deviation of $M_{\rm{Av}}$
from $M_{\rm{core}}$.

\section{Summary}

We mapped the submillimeter dust emission at 850 and 450\,$\mu$m and
the optical extinction of the IC\,5146 molecular cloud filament. All
data was smoothed to the resolution of the $A_V$ data, i.e. to
0.067\,pc ($30''$). The map covers a region of
$\sim2$\,pc$\times$0.3\,pc. Several peaks of $A_V>20$\,mag show up
along the east-west orientated dust ridge which we interpret as dense
prestellar cores.

\begin{itemize}
\item The maps of dust emission correspond well with the map of dust
  extinction. Most cores show up in all three maps. However, the map
  of dust extinction shows a higher contrast. The subsequent analysis
  relies on the estimated calibration accuracy, i.e. mainly on the
  assumption that the ratio of flux densities $S_{450}/S_{850}$ is
  accurate to within $\sim15$\,\%. This accuracy is established by
  accurate correction for the atmospheric opacity, by deconvolving the
  error beam at $450\,\mu$m, and by smoothing all three data sets to
  $30''$ resolution. The validity of the estimated calibration error
  is indicated by the smooth variation of dust temperatures over the
  surface of the dust ridge (Fig.\,4) and the small sampling errors of
  $T_{\rm{dust}}$ within the annuli of the radial core profiles
  (Fig.\,5). These errors are less than 1\,K for all annuli.
    
\item We constructed a map of dust temperature from the dust
  emissivities, assuming a constant dust emissivity index of $\beta=2$
  and only one dust temperature component along each line of sight.  A
  Monte Carlo simulation shows that the significant part of the dust
  temperature distribution is restricted to temperatures between
  $\sim12\,$K and $\sim20$\,K.  Cores have rather low temperatures of
  less than $\sim15\,$K, while the inter clump dust shows temperatures
  of about $\sim20\,$K.
  
\item A map of the dust absorption ratio
  $\kappa'=\kappa_{850}/\kappa_V$ was constructed by using in addition
  the map of optical extinctions. Its mean value corresponds well to
  the ratio found by \citet{mathis1990}.  The Monte Carlo simulation
  indicates a systematic and significant rise of $\kappa'$ with
  dropping dust temperature, i.e. from $\sim1.3\,10^{-5}$ at
  $\sim20$\,K to $\sim5\,10^{-5}$ at $\sim12$\,K.
  
  This translates into a corresponding variation of $\kappa_{850}$,
  assuming that $\kappa_{\rm{V}}$ stays constant.  Such a variation is
  in fact predicted by the grain evolutionary model of OH94 when
  coagulation and formation of ices on grain surfaces becomes
  important in the cold, dense cloud interiors.
  
  This is also consistent with the previously observed reduction of
  the CO gas phase abundance due to freeze out onto grains in the
  IC\,5146 center region (Paper II).

\item We identified four cores with high optical extinctions along the
  ridge.  Radial profiles of dust temperature show almost isothermal
  profiles for the inner radius of 0.067\,pc ($30''$). Two of the
  cores show a significant increase of dust temperature at larger
  radii which we attribute to the presence of externally heated dust of
  the inter clump medium.

\item Core masses $M_{\rm{Av}}$ derived from $A_V$ and integrated over
  a diameter of 0.13\,pc ($1'$) vary between 4 and 7\,\msol. Masses
  $M_{\rm{canon}}$ derived independently from the 850\,$\mu$m dust
  emissivity assuming canonical values for $\kappa_{850}$ and
  $T_{\rm{dust}}$ agree with $M_{\rm{Av}}$ by better than 30\%.  This
  indicates that both completely independent methods trace the
  distribution of total dust and gas column densities with that
  accuracy.  Masses $M_{\rm{core}}$ derived from $S_{850}$ taking into
  account the radial profiles of $T_{\rm{dust}}$ and $\kappa'$ are
  offset from $M_{\rm{Av}}$ by an almost constant factor of 0.85.

\end{itemize}

Dust fluxes at shorter wavelengths than $\sim450$\,$\mu$m are however
needed to better disentangle variations of dust temperature, dust
emissivity index $\beta$, and the absolute opacities.  This is because
the spectral energy distributions of cold cloud cores peak around
$\lambda\sim100 - 300\,\mu$m. Since this wavelength range is
inaccessible from the ground, it will be space-borne instruments like
PACS and SPIRE on board the Herschel space observatory, which is
scheduled to start in 2007, which will provide unprecedented
sensitivity and spatial resolution to access this wavelength regime
\citep[e.g.][]{andre2001garching}.

\begin{acknowledgements}
  We thank M.\,Walmsley and V.\,Ossenkopf for helpful comments on a
  first version. We are grateful to T.\,Huard for using a new
  NIR-dataset prior to publication, to derive optical extinctions in
  the area of the OFF beam. CJL acknowledges support from NASA Origins
  grant NAG5-9520. BM is supported by an Alexander von Humboldt
  research fellowship.
\end{acknowledgements}

\bibliographystyle{aa} 
\bibliography{aamnem99,kramer} 

\end{document}